\documentclass[oldversion,rnote]{aa}

\usepackage{graphicx}
\usepackage{txfonts}
\usepackage{natbib}
\usepackage{color}
\bibpunct{(}{)}{;}{a}{}{,} 

\newcommand{\qm}[1]{``#1''}

\newcommand{\be}{\begin{equation}}
\newcommand{\ee}{\end{equation}}

\def\,{\kern 0,25em}
\def\sss{\scriptscriptstyle}
\def\U{{\sss \!U}}
\def\L{{\sss \!L}}

\def\nuL{\nu_\L}
\def\nuU{\nu_\U}

\definecolor{gray}{rgb}{.6,.6,.6}
\definecolor{green}{rgb}{0,.6,0}
\definecolor{red}{rgb}{.9,0,0}
\newcommand*{\issue}[1]{\textcolor{red}{{\bf #1}}}
\def\tbd[#1]{\issue{{#1}\ldots\,(tbd)\,}}

\begin{document}
%
   \title{Resonant radii of kHz quasi-periodic oscillations\\
   in Keplerian discs orbiting neutron stars}
   \titlerunning{Resonant radii in discs orbiting neutron stars}

   \author{Z. Stuchl\'{\i}k
          \and
          A. Kotrlov\'a
          \and
          G. T\"{o}r\"{o}k
          }

   \institute{Institute of Physics, Faculty of Philosophy and Science,
            Silesian University in Opava,\\Bezru\v{c}ovo n\'{a}m. 13, CZ-74601
            Opava, Czech Republic\\
            \email{\,zdenek.stuchlik@fpf.slu.cz,\,andrea.kotrlova@fpf.slu.cz,\,terek@volny.cz}
            }

   \date{Received / Accepted }


  \abstract{Exchange of dominance between twin kHz quasi-periodic oscillations (QPOs) observed in some low-mass-X-ray-binaries (LMXB) suggests the possibility of a resonance between two oscillatory modes. We study the behaviour of the effective gravitational potential around specific resonant radii, and estimate the role of the higher-order terms governing the non-linear, anharmonic forcing. We discuss the impact it has on the mode amplitude in the linear and non--linear regimes. We also discuss a related possibility of lowering of the neutron star mass estimates from the highest observed QPO frequencies.}


   \maketitle
%

\section{Introduction}

High-frequency quasiperiodic oscillations (HF QPOs) of the X-ray brightness have been observed in several accreting binary systems containing neutron stars {\citep[NSs, see][]{Kli:2000:ARASTRA:,Kli:2006:CompStelX-Ray:,Bel-Men-Hom:2005:ASTRA:}} 
and black holes {\citep[BHs, see][]{McCli-Rem:2004:CompactX-Sources:,Rem-McCli:2006:ARASTRA:}}. In NSs they often appear in the form of two distinct modes (so-called \qm{twin peaks}) with frequencies correlated with the X-ray intensity and covering a relatively wide range $\sim 50-1300\,\mathrm{Hz}$. Henceforth, we adopt the convention of referring to twin-peak QPO modes as to lower and upper
QPOs and denote their frequencies as $\nu_{\mathrm{L}}$ and $\nu_{\mathrm{U}}$.  

The ratio of HF QPO frequencies when observed in black-hole systems in pairs is usually exactly, or almost exactly $3\!:\!2$, while in the NS systems the ratio $R\equiv\nuU/\nuL$ is concentrated around $R=3\!:\!2$ \citep[e.g.,][]{Tor:2005:ASTRN:,Abr-etal:2005:ASTRN:}. Most HF QPO models involve orbital motion in the inner regions of an accretion disc. Because of the $3:2$ ratio appearance, several proposed models consider resonances between the QPO modes \citep[][and others]{Abr-Klu:2001:ASTRA:,Abr-etal:2003:PUBASJ:,Lam-col:2003:arxiv:,Klu-etal:2004:ApJL:,Pet:2005:a:AA:,Pet:2005:b:AA:,Ter-Abr-Klu:2005:ASTRA:QPOresmodel,Sra-Tor-Abr:2006:ASTRA:,Stu-Sla-Tor:2007:,Stu-Kon-Mil-Hle:2008:ASTRA:GravExc,Stu-Kot:2009:GRG:,Muk:2009:ApJ:}.

\begin{figure*}[t]
  ~\hfill
  \resizebox{.98\hsize}{!}{\includegraphics{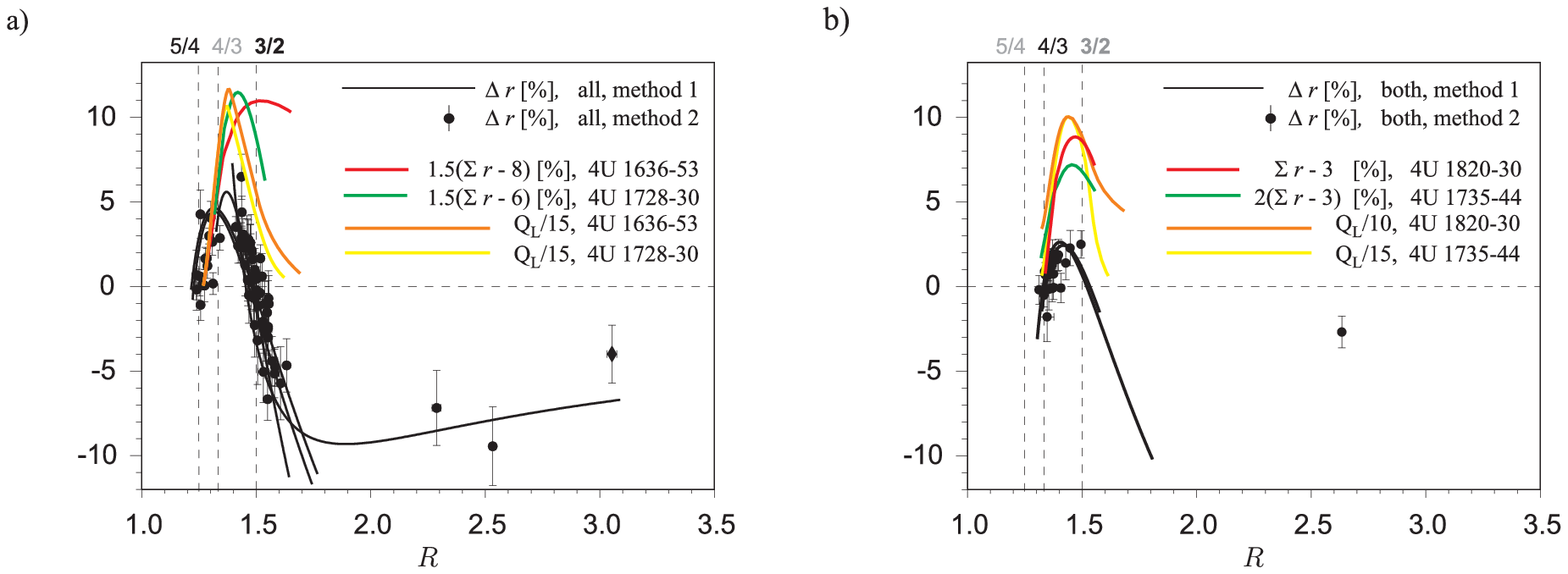}}
  \hfill~
  \caption{{Energy switch effect \citep[black datapoints and lines from][see this paper for details]{Tor:2009:AA:} and the twin QPOs {fadeaway} at low-frequency ratios. The colour-lines after \citet{Bar-Oli-Mil:2005:MONNR:,Bar-Oli-Mil:2006:ASTRN:DropCoher,Bar-Oli-Mil:2006:MONNR:QPO-NS}, \citet{men:2006}, and \citet{Tor:2009:AA:} indicate the drop in the behaviour of the total energy of the two QPOs (red and green lines) and the drop in the high cohererence of the lower QPO (orange and yellow lines). a) Frequency ratios emphasized in 4U 1636-53 and 4U 1608-52 (3:2 and 5:4) respectively 4U 0614+09 and 4U 1728-34 (3:2). b) 3:2 and 4:3 frequency ratios emphasized in the two sources 4U 1820-30 and 4U 1735-44.}}
\label{graf--Gabriel}
\end{figure*}

Quite recently \citet{Tor:2009:AA:} discovered that, in the six atoll NS systems, the rms amplitude difference $\Delta_r\equiv r_\mathrm{L}-r_\mathrm{U}$ is zero for the resonant frequency ratios $R=3\!:\!2$ at a frequency $\nu^{3:2}_{\mathrm{L}}$ specific to a particular system. The quantity $\Delta_r$ is positive for frequencies above $\nu^{3:2}_{\mathrm{L}}$ and negative for frequencies below $\nu^{3:2}_{\mathrm{L}}$.\footnote{It has been argued by \citet{Hor-etal:2009:AA:} that a natural explanation of such an \qm{energy switch} effect arises in the framework of the non-linear resonant phenomena.}  {Similar \qm{energy switch effect} also occurs at the frequencies corresponding to the resonant ratios $4\!:\!3$ or $5\!:\!4$.
Nevertheless, these ratios correspond to very high frequencies where kHz QPOs disappear. The \qm{disappearance} occurs since the QPO amplitudes and coherence times strongly decrease; see \citet{Bar-Oli-Mil:2005:MONNR:}, \citet{Bar-Oli-Mil:2006:ASTRN:DropCoher,Bar-Oli-Mil:2006:MONNR:QPO-NS}, \citet{men:2006}, \citet{Tor:2009:AA:}, \citet{Tor-etal:2008:ACTA:DistrKhZ4U1636-53,Tor-Bak-Stu-Cec:2008:ACTA:TwPk4U1636-53,Tor-etal:2008:ACTA:Clustering4U1636-53}, and \citet{Bou-Bar-Lin-Tor:2010:MNRAS:}. Figure~\ref{graf--Gabriel} illustrates both the energy switch effect and the QPO disappearance.}

We suggest that the inner edge of the disc can be related to the resonant radius corresponding to the $5\!:\!4$ {($4\!:\!3$)} frequency ratio. We study properties of the effective potential of geodesic motion in the vicinity of the resonant points corresponding to these frequency ratios. For identification of the QPO frequencies we assume a widely discussed \emph{relativistic precession model} (henceforth RP model) where $\nu_{\mathrm{L}} =\nu_{\mathrm{K}} - \nu_{\mathrm{r}}$ and  $\nu_{\mathrm{U}}= \nu_{\mathrm{K}}$ with $\nu_{\mathrm{K}}$ and $\nu_{\mathrm{r}}$ the Keplerian and radial epicyclic frequency of the geodesic motion \citep[][]{Ste-Vie:1998:ASTRJ2L:,Ste-Vie:1999:PHYRL:,Kar:1999:apj:,Kot-Stu-Ter:2008:CLAQG:}. 
This identification roughly merges with a model assuming m=-1 radial and m=-2 vertical disc-oscillation modes \citep[see discussion in][]{Tor-Stu-Bak:2007:CEURJP:,Tor-etal:2010:ASTRJ2:}.  We expect that our analysis is relevant not only to these two specific models but also to some others that assume resonances between the two QPO modes in the regions close to the innermost stable orbit and a frequency ratio decreasing with decreasing orbital radius.

We focus attention on the potential depth at the resonant radii and related maximal amplitude of the oscillations, and we determine the energy levels for the harmonic oscillations given by the expansion of the potential around the resonant radii (whereas the second order term of the expansion gives the frequency of the epicyclic motion). Furthermore, we estimate the role of the higher order expansion terms of the potential that govern the non-linear terms relevant to the resonant phenomena \citep{Nay-Moo:1979:NonOscilations:}. 

\section{Effective potential near the resonant radii at the Schwarzschild spacetime}

Using the dimensionless Schwarzschild radial coordinate $x = r/M$,
the effective potential of the radial motion in the
Schwarzschild spacetimes can be given in the form \citep{Mis-Tho-Whe:1973:Gra:}
\begin{equation}
V_{\mathrm{eff}}^2(x;L)=\left(1-\frac{2}{x}\right)\left(1+\frac{L^2}{x^2}\right),\quad L^{2}(x)  = \frac{x^2}{x-3},
\end{equation}
where $L$ is the specific angular momentum of a test
particle evaluated for the specific case of a circular orbit of given $r$ ($x$). 

\begin{figure}[t]
 \begin{center}
  \resizebox{0.9\hsize}{!}{\includegraphics{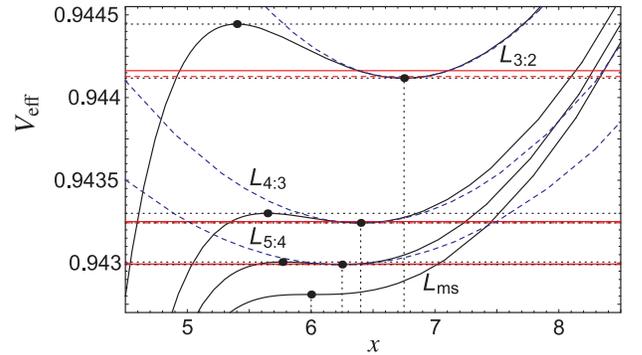}}
 \end{center} 
 \vspace{-2.5ex}
  \caption{The effective potentials $L_{\mathrm{n:m}}$ determined {from} the RP model (black lines). The blue dashed lines denote the
Taylor approximation up to the quadratic terms. The red solid lines correspond to the case of $A_\mathrm{p}/A_\mathrm{r} \sim 0.97$ and
the red dashed lines correspond to the case of $A_\mathrm{p}/A_\mathrm{r} = 0.99$.}
\label{figure-1}
\end{figure}

In the Schwarzschild spacetime, the orbital Keplerian frequency and
the radial epicyclic frequency read \citep[][]{ali-gal:1981,Ter-Stu:2005:ASTRA:} as
\begin{equation}
\nu_{\mathrm{K}} = \frac{F}{x^{3/2}}, \quad \nu_{\mathrm{r}} = \nu_{\mathrm{K}} \left(1 - \frac{6}{x}\right)^{1/2}, \quad  F =
\frac{\mathrm{c}^3}{2 \pi \mathrm{G}M}\,.
\end{equation}
The resonant radii for the RP model are determined by
\begin{equation}
\frac{\nu_{\mathrm{K}}}{\nu_{\mathrm{K}} - \nu_{\mathrm{r}}} =
\frac{n}{m}\label{RP-vztah}
~~\Rightarrow~~
x_{n:m} = \frac{6n^2}{m\left(2n-m\right)}.
\end{equation}
We thus find
$x_{3:2} = 6.75,~x_{4:3} = 6.40,~and x_{5:4} = 6.25$.
The relevant effective potentials are then given by
\begin{equation}
V_{\mathrm{eff}}(x;L)=V_{\mathrm{eff}}(x;L=L_{n:m}),\qquad
L^{2}_{n:m} = \frac{x^{2}_{n:m}}{x_{n:m} - 3}.
\end{equation}
These potentials can be expressed in the form
\begin{equation}
V_{\mathrm{eff}}(x;L_{n:m})=\left(1-\frac{2}{x}\right)\left(1+\frac{12n^4x^{-2}}{m(2n-m)(2n^2-2nm+m^2)}\right),
\end{equation}
and are presented for $n\!:\!m = 3\!:\!2$, $4\!:\!3$, $5\!:\!4$
in Fig.~\ref{figure-1}. Then we can determine the potential depth
at the resonant radii giving the energy barrier for the stability of the circular motion,
\begin{equation}
\Delta E_{n:m} = V_{\mathrm{max}}(x;L_{n:m}) -
V_{\mathrm{min}}(x;L_{n:m}).\label{en-rozdil-max-min}
\end{equation}
The results are given in Table~\ref{tabulka-1}. Similarly, we can determine the maximal amplitude $A_{n:m}$ allowed for the oscillations in the vicinity of the resonant radii (see again Table~\ref{tabulka-1}). We also give the ratio $\frac{A_{n:m}}{x_{n:m}}$, which demonstrates the maximal relative magnitude of the oscillations at the resonant radii. We can see that the potential depth and the maximal amplitude at the resonant radii strongly {decrease} with frequency ratio decreasing from $3\!:\!2$ down to $5\!:\!4$. Nevertheless, in all three cases, the maximal amplitude of the oscillations remains relatively large. The relative amplitude is largest for the $3\!:\!2$ ratio when $\frac{A_{n:m}}{x_{n:m}} \sim 0.34$. We note that the potential depth at $x_{5:4}$ is by more than one order ($\sim 1/20$) smaller than those at $x_{3:2}$, while the maximal amplitude of the oscillatory motion is smaller only by factor of $\sim 3$.

The $\mathrm{a_i}$ coefficients in Taylor expansions of the effective potential about the resonant radii are\\
for $x_{3:2}$:\\
$
  \mathrm{a_1}\!=\!0.944118,\,~\mathrm{a_2}\!=\!3.4444\!\times\! 10^{-4},\,~\mathrm{a_3}\!=\! 3.40148\!\times\! 10^{-5},\,~
  \mathrm{a_4}\!=\! -3.7857\!\times\!10^{-5},\, ~\mathrm{a_5}\!=\! 1.34255\!\times\! 10^{-5},\,~\mathrm{a_6}\!=\! -3.58131 \!\times\! 10^{-6},\,~
  \mathrm{a_7}\!=\! 8.32113 \!\times\! 10^{-7},\,~\mathrm{a_8}\!=\! -1.78353 \!\times\! 10^{-7}\,,\ldots\\
$
for $x_{4:3}$:\\
$
  \mathrm{a_1}\!=\!0.943242,\,~\mathrm{a_2}\!=\!2.37896 \!\times\! 10^{-4},\,~\mathrm{a_3}\!=\! 1.11514\!\times\! 10^{-4},\,~
  \mathrm{a_4}\!=-6.97262\!\times\! 10^{-5},\, ~\mathrm{a_5}\!=\! 2.35669\!\times\! 10^{-5},\,~\mathrm{a_6}\!=\! -6.36989 \!\times\! 10^{-6},\,~
  \mathrm{a_7}\!=\! 1.53105 \!\times\! 10^{-6},\,~\mathrm{a_8}\!=-3.43018\!\times\! 10^{-7}\,,\ldots\\
$
for $x_{5:4}$:\\
$
  \mathrm{a_1}\!=\!0.94299,\,~\mathrm{a_2}\!=1.67063 \!\times\! 10^{-4},\,~\mathrm{a_3}\!=\!1.60380 \!\times\! 10^{-4},\,~
  \mathrm{a_4}\!=\!-8.98277 \!\times\! 10^{-5},\, ~\mathrm{a_5}\!=\!3.00803\!\times\! 10^{-5},\,~\mathrm{a_6}\!=\!-8.20919 \!\times\! 10^{-6},\,~
  \mathrm{a_7}\!=\!2.00698 \!\times\! 10^{-6},\,~\mathrm{a_8}\!=-4.59198 \!\times\! 10^{-7}\,,\ldots\\
$

\begin{table}
\renewcommand{\arraystretch}{1.3}
\caption{The potential depth $\Delta E_{n:m}$ at the resonant
radii given by  relation~(\ref{en-rozdil-max-min}), the maximal
amplitude $A_{n:m}$ and the relative amplitude $A_{n:m}/{x_{n:m}}$
allowed for the oscillations in the vicinity of the resonant radii for the
RP model.}%
\label{tabulka-1}%
\centering
\begin{tabular}{|c|c c c|}
\hline
   $n\!:\!m$  & $3\!:\!2$ & $4\!:\!3$  & $5\!:\!4$
    \\
\hline
   $\Delta E_{n:m}$ & $3.27\times 10^{-4}$ & $5.77\times 10^{-5}$ & $1.51\times 10^{-5}$ \\
    $A_{n:m}$ & $2.31429$ & $1.21008$ & $0.75251$\\
    $A_{n:m}/{x_{n:m}}$ & $0.34286$  & $0.18908$ & $0.12040$\\
    \hline
\end{tabular}
\end{table}

\begin{table}
\renewcommand{\arraystretch}{1.3}
\caption{The potential depth $\Delta
E_{(V_{\mathrm{p}}-V_{\mathrm{MIN}})}$ and the related maximal
amplitude $A_\mathrm{p}$ of the harmonic oscillations for two
precision levels given by the ratios
$A_{\mathrm{p}}/A_{\mathrm{r}}=0.99$ and
$A_{\mathrm{p}}/A_{\mathrm{r}}=0.97$.}%
\label{tabulka-2}%
\centering
\begin{tabular}{|c|c c c|}
\hline
  $n\!:\!m$  & $3\!:\!2$ & $4\!:\!3$  & $5\!:\!4$
    \\
\hline \multicolumn{4}{|c|}{$\displaystyle
A_{\mathrm{p}}/A_{\mathrm{r}}\geq 0.99$}
\\
   \hline
   $\Delta E_{(V_{\mathrm{p}}-V_{\mathrm{MIN}})}$ & $9.5\times 10^{-6}$ & $3.9\times 10^{-7}$ & $6\times 10^{-8}$ \\
 $\displaystyle\frac{\Delta E_{n:m}}{\Delta E_{(V_{\mathrm{p}}-V_{\mathrm{MIN}})}}$ &
    $34.4$ & $148.1$ & $251.8$ \\
 $A_{\mathrm{p}}$ & $0.166085$ & $0.0404892$ & $0.0189513$ \\
 $A_{\mathrm{r}}$ & $0.167759$ & $0.0408933$ & $0.0191297$ \\
\hline\multicolumn{4}{|c|}{$\displaystyle
A_{\mathrm{p}}/A_{\mathrm{r}}\geq 0.97$}
\\
   \hline
   $\Delta E_{(V_{\mathrm{p}}-V_{\mathrm{MIN}})}$ & $45\times 10^{-6}$ & $8\times 10^{-6}$ & $3\times 10^{-6}$ \\
 $A_{\mathrm{p}}$ & $0.722944$ & $0.366759$ & $0.268010$ \\
 $A_{\mathrm{r}}$ & $0.729074$ & $0.370491$ & $0.272475$ \\
\hline
\end{tabular}
\end{table}

\noindent The expansion coefficients clearly demonstrate the importance of
the higher order terms. For the sake of completeness we give the expansion terms up to the eigth order, while for the non-linear resonant phenomena usually the third and fourth
terms are relevant \citep{Nay-Moo:1979:NonOscilations:,Stu-Kot-Tor:2008:ACTA:BHadmStrResPhen}.
We can see that at the $x_{3:2}$ radius the harmonic (second order) term is by one order higher in comparison with the third and fourth order terms. On the other hand, at the $x_{5:4}$ radius all
the second, third, and fourth order terms are of nearly the same magnitude, 
indicating a strong role of the non-linear resonant phenomena even
for small perturbations from the purely circular motion. At the
$x_{4:3}$ radius, all three terms are of comparable magnitude.

Now we can determine the regions that extend around the resonant radii where the oscillation can have a harmonic character, i.e., where only the second-order term of the Taylor expansion of the effective potential is relevant. We also determine the maximal energy level and the maximal amplitude of harmonic oscillations. We carry out the estimate for two precision levels given in terms of the ratio of the amplitude governed by the total effective potential ($A_\mathrm{r}$) and the amplitude implied by the second order (harmonic) part of its expansion ($A_\mathrm{p}$). The results are presented in Table~\ref{tabulka-2}. Clearly, both the energy level of the harmonic oscillations
and their amplitude strongly {decrease} with the resonant radius (and frequency ratio) decreasing. Table~\ref{tabulka-2} gives the evidence for an increasing role of the non-linear resonant phenomena {when the resonant radius decreases}. The harmonic character of the oscillations near $x_{5:4}$ is only possible for $\Delta E_{(V_{\mathrm{p}}-V_{\mathrm{MIN}})} \sim 6 \times 10^{-8}$, which is by factor $\sim 250$ smaller than the related potential depth.

\begin{figure}[t]
  ~\hfill\resizebox{.8\hsize}{!}{\includegraphics{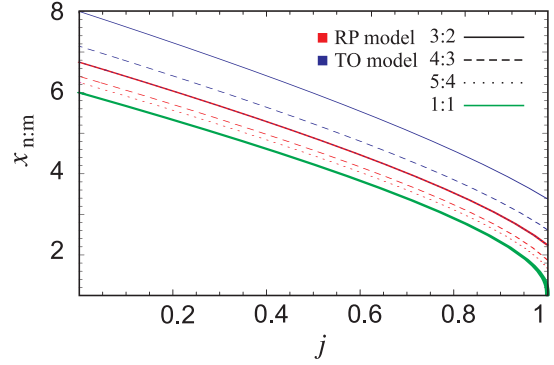}}\hfill~
  \caption{The $x_{n:m}\left(j;\,\mathrm{n,m}\right)$ functions for the RP model (red lines)  and the TO model (see Sect.~\ref{section:TO}). 
The thick green line represents the radius of the marginally stable orbit $x_{\mathrm{ms}}$ where $\nuU/\nuL=1$ for both models.}
  \label{figure-2}
\end{figure}

\section{Resonant radii at the Kerr geometry}

We next assume Kerr geometry describing the gravitational field around rotating BHs \citep{Bar-Pre-Teu:1972:ASTRJ2:}. This is in general a somewhat worse, but applicable, elegant, and fully analytical approximation to the case of the more complicated Hartle--Thorne geometry, which describes rotating NSs quite well \citep[][]{har-tho:1968,ber-etal:2005}. Moreover, for high NS masses, such as those implied by the RP model, the Kerr geometry approximation is very good \citep[][]{Tor-etal:2010:ASTRJ2:}.

Using the dimensionless Boyer--Lindquist radial coordinate $x =
r/M$ and the dimensionless spin $j$ the effective potential of the
radial motion in the Kerr spacetimes can be given in the form
\begin{eqnarray}
V_{\mathrm{eff}}(x;j;L)&&=\\
&&\frac{2 j L +
\sqrt{j^2-2x+x^2}\sqrt{x\left(2j^2+j^2x+xL^2+x^3\right)}}{x^3+j^2\left(2+x\right)}\nonumber,
\end{eqnarray}
where $L$ denotes the specific angular momentum of a test
particle. When the particle orbits in a
corotating circular orbit at given $r$ ($x$), its specific
angular momentum reads as
\begin{equation}
L(x;j)  =
\frac{x^2-2jx^{1/2}+j^2}{x^{1/2}\sqrt{x^2-3x+2jx^{1/2}}}.
\end{equation}
In the Kerr spacetime the orbital Keplerian frequency and the radial epicyclic frequency read as
\begin{equation}
\nu_{\mathrm{K}} = \frac{F}{x^{3/2}+j},\quad \nu_{\mathrm{r}} = \nu_{\mathrm{K}} \left(1 - \frac{6}{x} +
\frac{8j}{x^{3/2}} - \frac{3j^2}{x^2}\right)^{1/2}.
\end{equation}
Again, the zero point of the radial epicyclic frequency defines
the marginal stable geodesics $x_{\mathrm{ms}}$ that is implicitly
given by
\begin{equation}
j = j_{\mathrm{ms}}(x) \equiv \frac{4x^{1/2}}{3} - \frac{1}{3}
\sqrt{3x^2 - 2x}.
\end{equation}

\begin{figure*}
\centering
\resizebox{\hsize}{!}{\includegraphics{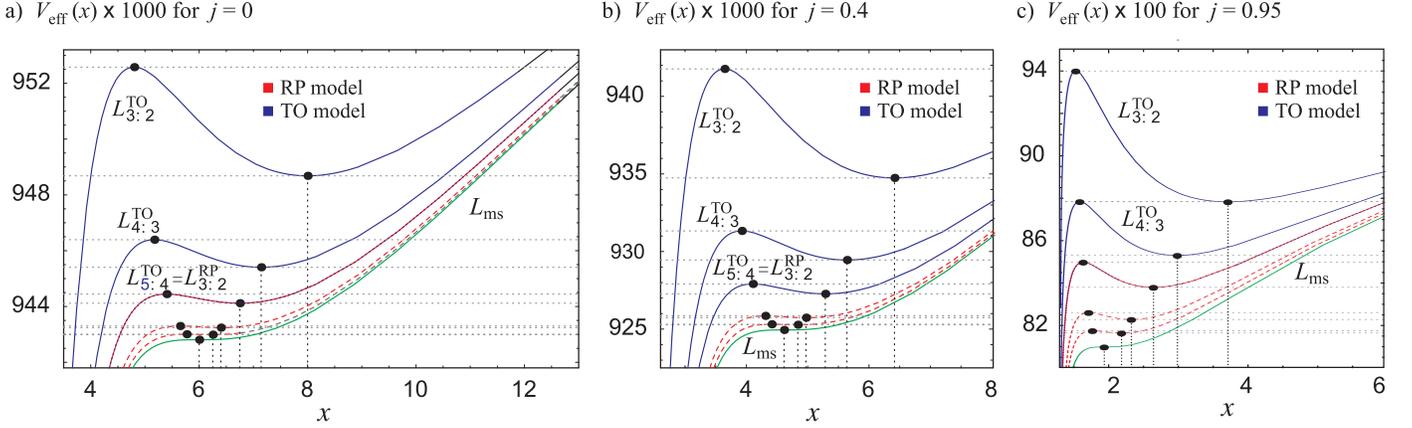}}
\caption{Effective potentials for the RP and TO models. The green curve $L_{\mathrm{ms}}$ corresponds to $R=1$ for both models. Also $\mbox{RP}_{3:2}\equiv \mbox{TO}_{5:4}$.} \label{figure-3}
\end{figure*}

\noindent The resonant radii for the RP model are now implicitly given by
\begin{equation}
j = j_{n:m}\left(x;n,m\right) \equiv
\frac{\sqrt{x}}{3} \left(4 - \sqrt{3x\left[1 -
\frac{(n-m)^2}{n^2}\right] - 2}\right)\label{vztah-pro-spin-Kerr}
\end{equation}
and presented in the explicit form $x_{n:m}(j)$ in Fig.~\ref{figure-2}. The angular momentum of test particles orbiting at the resonant radii is given by
\begin{eqnarray}
L_{n:m} = L_{n:m}(x_{n:m};j)  =
\frac{x_{n:m}^{2}(j)-2jx_{n:m}^{1/2}(j)+j^2}{x_{n:m}^{1/2}(j)\sqrt{x_{n:m}^{2}(j)-3x_{n:m}(j)+2jx_{n:m}^{1/2}(j)}}.\nonumber\\
\end{eqnarray}

\begin{figure*}
\resizebox{1\hsize}{!}{\includegraphics{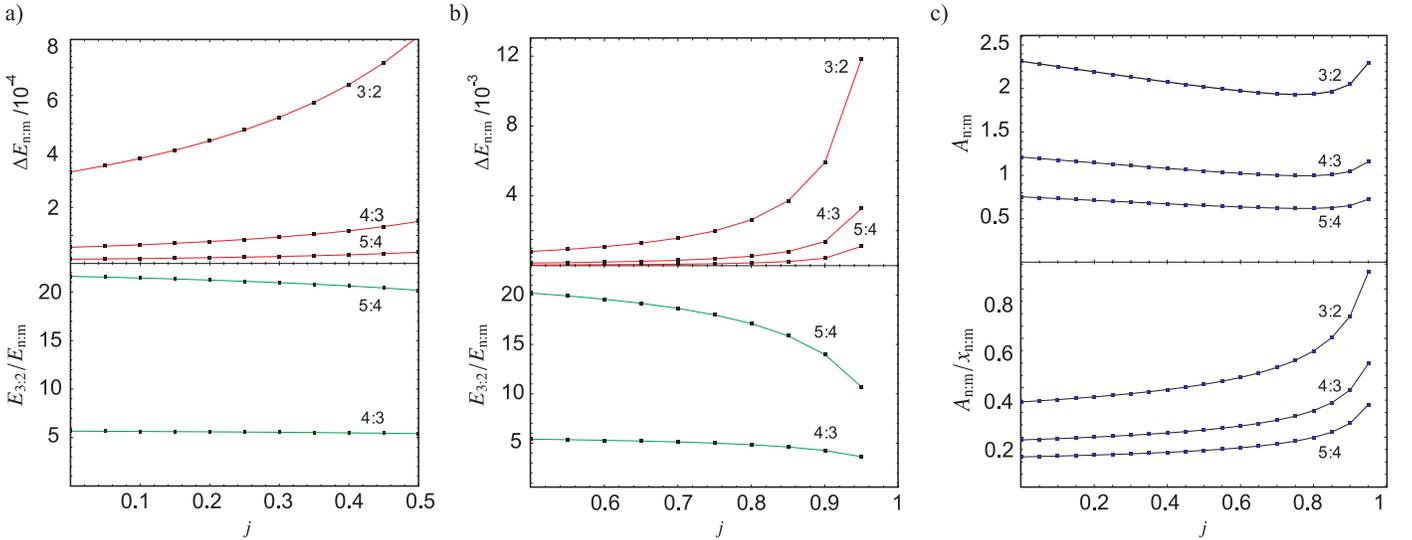}}
\caption{a) The spin $j$ profile for the potential depth giving the energy barrier of the stable circular motion at the resonant radii $\Delta E_{n:m}(j)$ for the RP model (top panel).  The relative potential barriers $\Delta E_{3:2}(j)/\Delta E_{n:m}(j)$ are illustrated for the considered resonant points (bottom panel). b) The same as a) but for high-spin geometries. c) The maximal amplitude $A_{n:m}(j)$ of oscillations about the resonant radii (top panel) and the ratio $\frac{A_{n:m}}{x_{n:m}}(j)$ for the RP model (bottom panel).} \label{figure-4}
\end{figure*}

\noindent The effective potential related to the angular momentum $L_{n:m}$
reads as
\begin{eqnarray}
&&V_{\mathrm{eff}}(x;j;L_{n:m}(x_{n:m};j))=\\
&&\frac{2 j L_{n:m} +
\sqrt{j^2-2x+x^2}\sqrt{x\left(2j^2+j^2x+xL_{n:m}^2+x^3\right)}}{x^3+j^2\left(2+x\right)}.\nonumber
\end{eqnarray}
For selected values of the spin ($j = 0, 0.4, 0.95$) the effective potential
with minimum at the resonant points is given in
Fig.~\ref{figure-3}. In Fig.~\ref{figure-4} we give the spin $j$ profile for the
potential depth at the resonant radii $\Delta E_{n:m}(j)$, the related maximal amplitude $A_{n:m}(j)$ of oscillations about the resonant radii, and the ratio $\frac{A_{n:m}}{x_{n:m}}(j)$.

We can see that the potential depth $\Delta E_{n:m}(j)$ and the
ratio $\frac{A_{n:m}}{x_{n:m}}(j)$ with growing $j$ increase.
This increase is most restricted for the resonant radius $x_{5:4}$
and is greatest for $x_{3:2}$. The amplitude of the oscillations
$A_{n:m}$ {decreases} when $j$ grows up to $\sim 0.9$, and then it
increases. On the other hand, the relative magnitude of the
oscillations $\frac{A_{n:m}}{x_{n:m}}$ grows within the whole allowed
interval of $j$. This indicates that the potential detectability of
the resonant oscillations grows with increasing $j$. Especially near $x_{3:2}$ the relative amplitude is very large, $>0.34$.


\section{Discussion and conclusions}

{The role of the non-linear terms determining the non-linear resonant phenomena grows for the decreasing resonant radius correponding for the RP model to a decreasing frequency ratio. Extension of the possible harmonic radial oscillations decreases with decreasing resonant radius. This decline is strongest at $x_{5:4}$, reaching almost two orders, which is much higher compared to those at $x_{3:2}$ where it is higher by only one order.  The expected magnitude of oscillations around the resonant radii can then be quite large when the non-linear regime is entered, and it can be of the same order as the resonant radius. We note that the observability of such phenomena is expected even for oscillations with the amplitude by one order smaller than $x_{n:m}$ {\citep[][]{bur:2005:apj,Sch-Rez:2006:ASTRJ2:QPOsRelTor,Abr-etal:2007,Bur:2008}}.

\begin{figure}[t!]
  ~\hfill\resizebox{0.8\hsize}{!}{\includegraphics{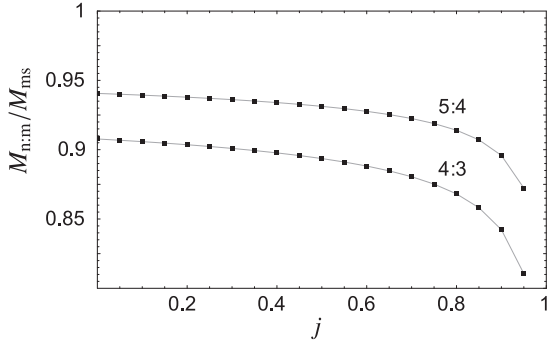}}\hfill~
  \caption{Lowering of the estimated neutron-star mass.}
  \label{figure-mass}
\end{figure}

If the edge of the accretion disc is located at the resonant radius
$x_{n:m}$ (with $n\!:\!m = 4\!:\!3$  or $5\!:\!4$) rather than at
the marginally stable orbit $x_{\mathrm{ms}}$, the maximal
frequency $\nu_{\mathrm{max(o)}}$ observed in a given source has
to be attributed to the Keplerian frequency
$\nu_{\mathrm{K}}(x_{n:m})$ rather than to
$\nu_{\mathrm{K}}(x_{\mathrm{ms}})$, which is {often applied in
QPO-ISCO-estimates for the mass of the NS \citep[e.g.][]{Bar-Oli-Mil:2005:MONNR:,Bar-Oli-Mil:2006:ASTRN:DropCoher,Bar-Oli-Mil:2006:MONNR:QPO-NS}}. Therefore, we
expect lowering of the NS mass estimated {from the highest QPO frequencies} owing to the
shift to higher radii of the orbital motion. Since 
$\nu_{\mathrm{max}} = \nu_{\mathrm{K}}(x_{\mathrm{ms}}(M_{\mathrm{ms}})) = \nu_{\mathrm{K}}(x_{n:m}(M_{n:m}))$, 
we immediately arrive at the relation
\begin{equation}
            \frac{M_{n:m}}{M_{\mathrm{ms}}} = \frac{x_{\mathrm{ms}}^{3/2} + j}{x_{n:m}^{3/2} +
            j}\,,\label{vztah-hmotnost}
\end{equation}
which determines the lowering of the estimated NS mass (see
Fig.~\ref{figure-mass}). Assuming the Schwarzschild geometry
($j=0$), we find the relations
\begin{equation}
            M_{5:4} = 0.9406\, M_{\mathrm{ms}},\quad M_{4:3} = 0.9077\, M_{\mathrm{ms}} .
\end{equation}
The growing spin causes a further shift, and at the edge of the spin interval for the slowly
rotating NS we find
\begin{equation}
            M_{5:4}(j \sim 0.4) = 0.9340\, M_{\mathrm{ms}},~M_{4:3} (j \sim 0.4) = 0.8977 M_{\mathrm{ms}}.
\end{equation}
Such corrections are relevant and tend to lower the estimated masses down to the canonical value
$M \sim 1.4\,M_\odot$.

\label{section:TO}

{Throughout} the paper we have assumed the RP model frequency relations. For comparison we discuss the behaviour of the effective
potential in the Schwarzschild spacetime at the resonant radii relevant for an alternative kHz QPO model, namely the model of deformed disc oscillations trapped in the gravitational field of the compact object that was introduced by \citet{Kato:2009:PASJ:} (henceforth the TO model). In the TO model the
resonance condition reads
 ${(2\nu_{\mathrm{K}}-\nu_{\mathrm{r}}})/2{\left(\nu_{\mathrm{K}}-\nu_{\mathrm{r}}\right)}={n}/{m}$. 
We show the effective potentials with minima at the resonant radii of the TO model for the Schwarzschild spacetime, and for selected values of spin $j$ in Fig.~\ref{figure-3}. For the Schwarzschild spacetime we give the resonant radii, the potential depth, and the maximal amplitude of the oscillations at these radii in Table~\ref{tabulka-TO}. Notice that $x_{3:2}^{\mathrm{RP}} = x_{5:4}^{\mathrm{TO}}$. Therefore, for
all spin values, the resonant radii, potential depth and maximal amplitude corresponding to the TO model are greater than those related to the RP model.

\begin{acknowledgements}
This work was supported by the Czech grants MSM~4781305903,
LC06014, and GA\v{C}R 202/09/0772. {The authors also acknowledge the support of the internal grants of the Silesian University in Opava, FPF SGS/1/2010 and FPF SGS/2/2010.} One of the authors (ZS) would like
to express his gratitude to the Committee for Collaboration of the Czech Republic with CERN for the support, and the Theory Division of CERN for the perfect hospitality. 
\end{acknowledgements}

\begin{table}
\renewcommand{\arraystretch}{1.3}
\caption{Quantities $\Delta E_{n:m}$, $A_{n:m}$, and $A_{n:m}/{x_{n:m}}$ for the TO model.}%
\label{tabulka-TO}%
\centering
\begin{tabular}{|c|ccc|}
\hline
   $n\!:\!m$  & $3\!:\!2$ & $4\!:\!3$  & $5\!:\!4$
    \\
\hline
   $\Delta E_{n:m}$ & $3.9\times 10^{-3}$ & $9.88 \times 10^{-4}$ & $3.27\times 10^{-4}$ \\
    $A_{n:m}$ & $7.2$ & $3.65112$ & $2.31429$\\
$A_{n:m}/{x_{n:m}}$ & $0.9$  & $0.51116$ & $0.34286$ \\
\hline
\end{tabular}
\end{table}


\end{document}